\documentclass[preprint,authoryear,12pt]{elsarticle}

\usepackage{graphicx}

\usepackage{amssymb}

\usepackage[
a4paper=true,%
breaklinks=true,%
colorlinks=true,%
pdfauthor={Nindos et al.},%
pdftitle={Solar Physics with the Square Kilometre Array}%
]{hyperref}

\usepackage{booktabs}
\usepackage{threeparttable}

\journal{Advances in Space Research}

\newcommand{\arcmin}{{\hbox{$^\prime$}}}
\newcommand{\arcsec}{{\hbox{$^{\prime\prime}$}}}
\newcommand{\degr}{{\hbox{$^\circ$}}}

\begin{document}

\begin{frontmatter}



\title{Solar Physics with the Square Kilometre Array}


\author{A. Nindos\corref{mycorrespondingauthor}}
\address{Physics Department, University of Ioannina, GR-45110 Ioannina, Greece}
\cortext[mycorrespondingauthor]{Corresponding author}
\ead{anindos@uoi.gr}



\author{E.P. Kontar\corref{}}
\address{School of Physics \& Astronomy, University of Glasgow, G12, 8QQ,
Glasgow, UK}
\ead{Eduard.Kontar@glasgow.ac.uk}

\author{D. Oberoi}
\address{National Centre for Radio Astrophysics, Tata Institute of Fundamental
Research, Pune 411007, India}
\ead{div@ncra.tifr.res.in}

\begin{abstract}
The Square Kilometre Array (SKA) will be the largest radio telescope
ever built, aiming to provide collecting area larger than 1 km$^2$.
The SKA will have two independent instruments, SKA-LOW comprising of
dipoles organized as aperture arrays in Australia and SKA-MID
comprising of dishes in South Africa. Currently the phase-1 of SKA,
referred to as SKA1, is in its late design stage and construction is
expected to start in 2020. Both SKA1-LOW (frequency range of 50-350
MHz) and SKA1-MID Bands 1, 2, and 5 (frequency ranges of 350-1050,
950-1760, and 4600-15300 MHz, respectively) are important for solar
observations. In this paper we present SKA's unique capabilities in
terms of spatial, spectral, and temporal resolution, as well as
sensitivity and show that they have the potential to provide major new
insights in solar physics topics of capital importance including (i)
the structure and evolution of the solar corona, (ii) coronal heating,
(iii) solar flare dynamics including particle  acceleration and
transport, (iv) the dynamics and structure of coronal mass ejections,
and (v) the solar aspects of space weather.  Observations of the Sun
jointly with the new generation of ground-based and space-borne
instruments promise unprecedented discoveries.
\end{abstract}

\begin{keyword}
Sun \sep Sun:radio radiation \sep Sun:corona \sep Sun:flares 
\sep Sun:coronal mass ejections
\end{keyword}

\end{frontmatter}

\parindent=0.5 cm

\section{Introduction}

Although solar physics is one of the most mature branches of
astrophysics, the Sun confronts us with a large number of outstanding
problems that are fundamental in nature.  These problems include the
determination of the structure and dynamics of the solar atmosphere,
the magnetic field evolution in the chromosphere and corona, coronal
heating, the physics of impulsive energy release, energetic particle
acceleration and transport, the physics of coronal mass ejections
(CMEs) and shocks, as well as the solar origin of space weather
drivers.

Solar radio astronomy has already provided important insights into
these problems because (i) two of the natural frequencies of the solar
magnetized plasma, the electron plasma frequency, $\nu_{pe}$, and the
electron gyrofrequency, $\nu_{ce}$, fall into the radio band, and (ii)
radio emission is sensitive not only to the properties of radiating
electrons but also to the properties of the ambient plasma and magnetic field.
However, the full exploitation of the diagnostic potential of solar
radio emission is instrumentally demanding because solar radio sources
can be both small 
\citep[$\lesssim$ 1$\arcsec$ in localized episodes of energy  release; e.g.
see the VLBI observations by][]{Tapping83}
and large 
\citep[$\gtrsim$ 1$\degr$ in radio CMEs; e.g.][]{Bastian01}, weak
($<1$~sfu in the quiet Sun) and strong ($>$ 10$^5$~sfu in large
flares). Furthermore flux variations down to msec time scales and
narrow-band features of relative bandwidth down to 1-0.01\% are often
observed in coherent flare/CME-related bursts.  Therefore,
non-incremental progress in solar radio astronomy requires
high-cadence, high-sensitivity, large dynamic range,  snapshot imaging
interferometric observations over a large bandwidth with adequate
$u-v$ coverage and angular and frequency resolution.  Unfortunately,
there has been no radio instrument that is capable of meeting all of
the above requirements.

Historically, solar radio observations have been performed along two
lines: spectroscopy and imaging observations.
Spectroscopic observations (primarily at decimeter, meter, and
decameter wavelengths) provide spatially unresolved data with time and
frequency  resolution of about $\lesssim$ 1 s and $\lesssim$ 1 MHz,
respectively, while  interferometric imaging observations take place
at discrete frequencies providing data with angular resolution from a
few arcsec to a few arcmin depending on the frequency of
observations.

During the last couple of decades, old solar-dedicated and 
general-purpose radio
interferometers have been upgraded and some of them are now 
capable of performing
solar radio spectroscopic imaging and the same is/will be true for new
instruments (either already deployed or under construction).
The solar-dedicated instruments include the upgraded
Expanded Owens Valley Solar Array \citep[EOVSA; see][]{Gary12}, the
Mingantu Ultrawide Spectral Radioheliograph \citep[MUSER;
see][]{Yan16}, the Nan\c{c}ay Radioheliograph 
\citep[NRH; see][]{Kerdraon97}, the Nobeyama Radioheliograph 
\citep[NoRH; see][]{Nakajima94}, and the upgraded Siberian Radioheliograph
\citep[Siberian RH; see][]{Lesovoi14} while the general-purpose
instruments include the expanded Karl G. Jansky Very Large  Array
\citep[VLA; see][]{Perley11}, the Low Frequency Array \citep[LOFAR,
e.g.][]{2013A&A...556A...2V}, the Murchison  Widefield Array \citep[MWA,
see][]{Tingay13,Bowman13}, the Long Wavelength Array
\citep[LWA; see][]{Ellingson09}, and the Giant Metre-wave Radio Telescope
\citep[GMRT; see][]{Swarup91}. A summary  of the specifications of
these instruments (in alphabetical order) is given in Table 1.

\begin{table}
\centering
\begin{threeparttable}
\caption{Instruments capable of performing solar radio spectroscopic imaging}
\begin{tabular}{llllll}
\hline
Instrument & Frequency & Spectral &  Time   & Angular & Solar \\
           & Range     & Resolution & Resolution  & Resolution & dedicated \\
           & (GHz)       & (MHz) & (msec)             & (\arcsec) &  \\
\hline
EOVSA       & 1-18           & 50                   & 20         & 3-57 & Yes \\
GMRT        & 0.15-1.50  & 0.05  & 100 & 2-20 & No \\
LOFAR       & 0.03-0.24   & 0.1         & 10  & 60-540  & No \\
LWA         & 0.02-0.08    & 0.008        & 1    & 2-8\tnote{a} & No \\
MUSER       & 0.4-15         & 25                 & 25-200     & 1.3-50 & Yes \\
NoRH  & 17, 34      & 1700      & 100  & 6-12  & Yes \\
NRH & 0.15-0.45   & 23-48     & 250  & 18-240 & Yes \\
MWA         & 0.08-0.30   & 0.04      & 500  & 16-60 & No \\
Siberian RH & 4-8            & 10                   & 560  & 15-30 & Yes \\
VLA         & 1-50       &   1         & 100 & 1-35 & No \\
\hline
\end{tabular}
\begin{tablenotes}
\item[a]{Currently only two LWA stations have been deployed; the angular 
resolution cited here refers to the originally envisaged array.}
\end{tablenotes}
\label{table1}
\end{threeparttable}
\end{table}

The Square Kilometre Array (SKA) will be a general-purpose radio
instrument  whose construction is expected to start in 2020 and whose
characteristics will significantly exceed the capabilities of the instruments
of Table 1.  It will be capable of
performing solar  observations which will have the potential to
transform solar radio astronomy in particular, and solar physics in
general. A testament to this argument is the important results that
have come out of data obtained with SKA's ``precusor'' and ``pathfinder'' 
projects: the
LOFAR \citep[e.g][]{Morosan14,Morosan15,Morosan17,Reid17,Kontar17,2018ApJ...856...73C},
the MWA \citep[e.g.][]{Oberoi11,Mohan17,McCauley17,Suresh17,Cairns18,Sharma18},
the VLA \citep[][]{2009IAUS..257..529S,Chen13,2014ApJ...784...68K,Chen15},
and the LWA \citep[][]{Tun15}.

This paper is organized as follows.   In Section 2, we outline the
major features of the SKA and in Section 3 we discuss the
particulars of solar observations with it.  In Section 4 we
briefly outline solar radio emission mechanisms, and in Sections 5-9
we discuss open issues on solar radio astronomy and how the relevant
SKA observations can bring results of transformative nature. In
Section 10 we outline possible synergies between the SKA and other
instruments, and in Section 11 we summarize the paper.  Note that in
this paper we restrict ourselves only to the discussion of solar
physics problems that could be addressed by SKA observations; a discussion
of heliospheric physics problems which will be addressed by the SKA
lies outside the scope of this paper. A broader but shorter review discussing 
SKA's potential impact on both solar and heliospheric physics is available 
in \citet{Nakariakov15}.

\section{Overview of the instrument}

\subsection{Basic features}

This section is based on information from various documents that can
be found at \url{https://astronomers.skatelescope.org}.  The SKA is an
international project to build the world's largest and most sensitive
radio telescope. When completed the collecting area of the instrument
will be about one square kilometer, hence its name. The instrument
will be built in two phases, known as SKA1 and SKA2.  SKA1 will
correspond to about 10\% of the final collecting area and its
deployment will start in 2020 while commissioning activities are
expected to start in 2024. SKA2 will correspond to the full final
system and its construction will start,
subject to the performance of SKA1, after 2030.

\begin{figure}
\centering
\includegraphics[scale=0.5]{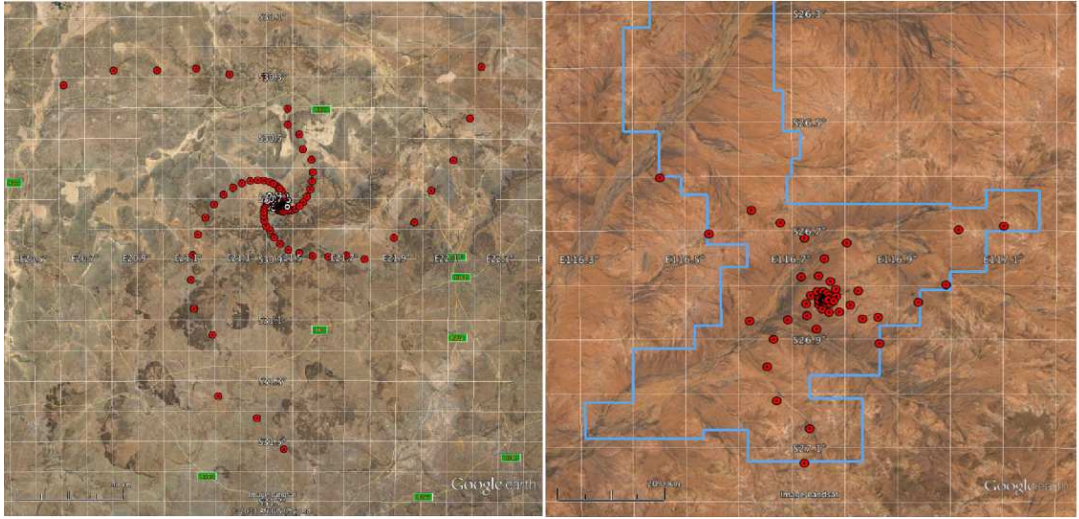}
\caption{Geophysical maps with the SKA1-MID (left) and SKA1-LOW (right)
configurations. The horizontal lines at the bottom left corners of the
maps correspond to 40/20 km (left/right map). Image credit: SKAO.}
\label{fig1}
\end{figure}

The SKA1 will consist of two arrays, SKA1-LOW and SKA1-MID, that will
be built in Australia and South Africa, respectively.  The expected
configuration  of the two arrays is presented in the geophysical maps
of Figure \ref{fig1}.  The maximum baselines of the arrays is expected
to be about 65 km for SKA1-LOW and about 150 km for SKA1-MID.

\begin{figure}[h]
\centering
\includegraphics[width=\textwidth]{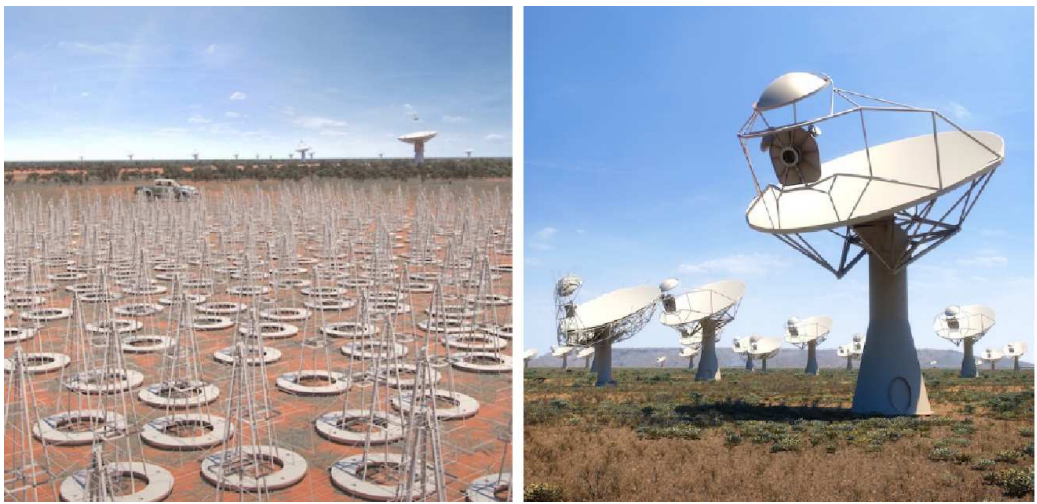}
\caption{Artist's impression of the SKA1-LOW aperture arrays (left)
and SKA1-MID dishes (right). Image credit: SKAO.}
\label{fig2}
\end{figure}

The SKA1-LOW will observe from $\sim$50 to 350 MHz and it will be
deployed in the Murchison desert of Western Australia (central
coordinate 26$\degr$ 41$\arcmin$ 49$\arcsec$S 116$\degr$ 37$\arcmin$
53$\arcsec$E) in the same region as the Australian SKA Pathfinder
(ASKAP)  and MWA arrays.  It will include about 131000 simple antennas
(e.g. log-periodic dual-polarization dipole elements) that are
arranged in 100-m-diameter stations  each hosting 90 elements.  In
each station the signal of all elements will be  added together
electronically, in phase, allowing the formation of an ``aperture
array'' (see Figure~\ref{fig2}, left).  The separation 
between stations will increase from the central part of the array
toward its outer edge, reaching several kilometers there.

The SKA1-MID will observe in the range from 350 MHz to 15.3 GHz which
will be divided into three frequency bands (Band 1: 0.35-1.05
GHz, Band 2: 0.95-1.76 GHz, and Band 5: 4.60-15.30
 GHz). The array will include 133 15-m-diameter dishes (see
Figure~\ref{fig2}, right) and will also incorporate the 64
13.5-m-diameter  dishes of the MeerKAT array. SKA1-MID will be
deployed in the Karoo desert of South Africa, ~500 km north of Cape
Town (central coordinate 30$\degr$ 43$\arcmin$ 16$\arcsec$S 21$\degr$
24$\arcmin$ 40$\arcsec$E). The dishes will be placed along a
three-armed spiral (see Figure~\ref{fig1}, left); along each arm the
separation of dishes will increase with the  distance from the center
of the array.

\begin{table}
\caption{Summary of SKA1 specifications}
\begin{tabular}{lll}
\hline
Parameter & SKA1-LOW value  & SKA1-MID value \\
\hline
Frequency range (MHz)   & 50-350 & 350-15300   \\
Angular resolution ($\arcsec$) & 4-24 &  0.025-1 \\
Time resolution (s) & 0.001-9 & 0.001-9 \\
Frequency resolution (kHz) & 5 & 15-76 \\
Field of view (deg$^2$) & 2-39 &  $\leq$ 1.4  \\
Largest angular structure ($\arcmin$) & 60-500 & 15-80  \\
Sensitivity ($\times 10^{-3}$ sfu) & 0.46-4.00  & 0.16-0.90  \\
\hline
\end{tabular}
\label{table2}
\end{table}

In Table 2 we present a summary  of SKA1's specifications. They
represent an unprecedented improvement over all existing
radio telescopes and will allow SKA1 to contribute to decisive
advances in practically all branches of modern astrophysics
\citep[e.g. see][and references therein]{Braun15}.

The SKA2 detailed specifications will be finalized depending on the
progress of SKA1. It is anticipated that both the SKA2-LOW
and -MID arrays  will form five 300-km-long spirals. There are plans
for the deployment of a few extra dishes to southern and central
African states (i.e. away from the instrument's core spiral configuration)
that will yield maximum baseline of about 3000 km.

The development of an instrument of such size poses difficult
construction and engineering problems. Furthermore, even the SKA1 will
produce data rates of several TB/s whose processing will require
central supercomputers capable of in excess of 100 petaflops
(i.e. 10$^{17}$ floating point operations per second) of  raw
processing power. The technical consortia  that are responsible to
build the SKA rely on the  recent huge progress both in the design of
antennas and wide-bandwidth feeds and in the transport and process of
large amounts of data. These recent developments have resulted in the
development of instruments (either new or upgraded old ones) that are
engaged in SKA-related technology and science  studies. These
instruments are called ``SKA precursors'' (if located at future SKA sites,
e.g. the MWA) or ``SKA pathfinders'', the latter include the VLA, the
LOFAR, the GMRT and others (for a complete census, see
\url{https://tinyurl.com/yaageuhz}).

\subsection{Modes of observations and data analysis}

SKA is an evolving project and several of its engineering, hardware 
and software aspects have not been finalized yet. The technical aspects of 
the SKA project (including software) are outlined in \citet{Hall05} while 
more up-to-date information can be found at 
\url{https://astronomers.skatelescope.org}.

The signals collected from MID and LOW arrays will be digitized
and fed into the instument's
``Central Signal Processor'' (CSP)\footnote{For more details see
 \url{https://www.skatelescope.org/csp}.}. CSP's main task will be
the collection, correlation, filtering, and analysis of the
observational data, according to the requirements of each observing
run.  To meet the requirements imposed by the SKA's specifications,
the CSP will utilize extremely fast computers (see Section
2.1). Furthermore, to reduce power consumption, most of the signal
processing will be done in Field-Programmable Gate Arrays (FPGAs) and
Gaphics Processing Units (GPUs).

Telescope resources will be enhanced by allowing users of both
telescopes (MID and LOW) to divide  collecting area into up to 16
subarrays and operate each subarray as an independent
instrument. Users will be allowed to schedule their observing run,
select observing mode, and determine the timeline of their
observations for each subarray independently.

The SKA will conduct primarily two kinds of observations,
interferometric  imaging and beamforming. All interferometric imaging
observations will, by definition, be spectroscopic. For a given
subarray operating in ``interferometric mode'', each pair of antennas
will be cross-correlated to provide full-polarization visibilities
across the requested bandwidth and number of channels.

In the ``beamforming mode'' each subarray can form several tied-array
beams and process data for each beam independently:

\begin{itemize}

\item SKA1-MID will be able to form up to 1500 ``Pulsar Search'' beams, 
spread over up to 16 subarrays, each covering 300 MHz, based on the 
selected antennas within 10 km of the subarray center. Similar operations
are anticipated for SKA1-LOW for up to 500 beams.

\item Both MID and LOW telescopes will be able to form up to 16 ``Pulsar
Timing'' beams. The setup will be similar to the ``Pulsar Search'' one, with
the exception that each beam will cover full input bandwidth for the observing
band.

\end{itemize}

The output of the CSP will be forwarded to the ``Science Data
Processor'' (SDP)\footnote{For more details see
\url{https://www.skatelescope.org/sdp}.}  for further reduction and
post-processing in order to derive scientifically useful results.  The
objective of the SDP subsystem is to build the necessary data analysis
software and pipelines, and the necessary hardware platform on which
these ``realtime'' or pseudo-realtime pipelines will run. The details
of the SDP data products are still being worked out but it is
anticipated that at least calibrated visibilities and
continuum and/or spectral-line image cubes will be made available to SKA
scientists.  The SDP operations will naturally take place close to the
telescope sites  and the results will be stored somewhere not too far
away. There is a  plan/desire to have SKA data centers set up across
the world, which will host copies (or subsets) of the SKA data
archives, where scientists can access the processed SKA data
products. These centers are expected to be places where the vast
majority of the scientists will interact with and analyze the SKA
data. Formally, they lie outside the scope of the SKA project and are
being financed independently by the SKA member countries.

SKA is expected to have a set of Key Science Projects (KSPs)
(typically with observing requirements of order 1000 hours or more)
and also ``Principal Investigator'' (PI)-driven projects. The KSPs are
still being formulated and the division of observing time between KSPs
and PI-driven projects is unclear. SKA will also have the requirement
to distribute observing time roughly in proportion to the
contributions by the member countries, although final decisions have not
been reached yet.

\section{Solar observations with the SKA1}

Scientists interested in using the SKA for their research have formed
``Science Working Groups'' (SWGs) which are advisory groups that
provide input to the SKA Organization (SKAO) ``...on the design,
commissioning, and future operations of the SKA that are likely to
affect the Observatory's scientific capability, productivity, and user
relations''\footnote{Excerpt taken from the Terms of Reference for
SKA's SWGs (see \url{https://tinyurl.com/ycmtjdrt}).}

One of them is the ``Solar, Heliospheric and Ionospheric'' (SHI) SWG\footnote{https://www.skatelescope.org/shi/}.
It  has more than 60 members from four continents and 20 countries and
it is currently chaired by E.P. Kontar (Glasgow) and  D. Oberoi (Pune).
The scientific interests  of the SHI SWG include the quiet Sun,
non-flaring active regions (ARs),  solar flares, CMEs, the  solar
wind, the Sun-Earth system, and the ionosphere.

The SHI group investigates the feasibility, hardware-wise, of solar
observations with both SKA1-LOW and -MID.  The SKA will be
able to observe the Sun both in interferometric imaging and beamforming
modes (see Section 2.2).

Low-frequency solar radio burst observations are in many ways similar
to observations of bright astrophysical radio sources and many general
purpose techniques can be ported to solar observations.  As for the
SKA1-MID, it has been established that: (1) the dishes can be pointed
to the Sun without raising the temperature of the receiver box by an
unacceptable amount, and (2) the low-noise amplifiers  themselves are
not expected to be saturated by the slowly-varying solar radio
emission or even solar radio bursts because they have been designed to
have sufficient  head-room to be able to deal with the occasional
presence of strong radio-frequency interference. On the other hand,
for the digital sampling of  the data  there will be a need to adjust
the gains/attenuations at appropriate  stages in the signal chain to
ensure that the sampling levels chosen are  appropriate for the range
of the voltages spanned by the solar signal.

Another important issue is that simultaneous observations with the LOW
and  MID arrays (especially of transient phenomena) will be very
difficult due to the large separation of the two arrays.  Simultaneous
observations at the three SKA1-MID Bands (see Section 2) will be
possible, in principle. This will, however, require configuring the
array as three independent sub-arrays operating at each of the three
bands.

Beamforming observations of the Sun will be possible by using the
pulsar setup briefly described in Section 2.2. The observing requirements
of the pulsar setup in terms of time and frequency resolution will be
more extreme than those needed for solar observing. The large number of
beams that will be provided simultaneously will be sufficient to tile the
SKA1-MID beam and the whole Sun in the case of SKA1-LOW.

As mentioned in Section 2.2, the development of the SDP is still 
in its early phases. Details of the data products it will eventually
deliver are under discussion and the computational requirements to
meet them are being estimated. No software tailored to the unique
snapshot spectroscopic imaging requirements of solar observations has
yet been developed either by the SKA project or by the SHI SWG. However,
the solar and heliospheric science team of the MWA, a SKA precursor,
has recently developed an automated interferometric solar imaging
pipeline, a first step towards building experience and understanding
the issues associated with solar imaging from this class of
instruments (Mondal et al., in preparation).

\section{Solar radio emission mechanisms}

The electrons that produce solar radio emission can be either thermal
or nonthermal and the emission mechanisms either coherent or
incoherent 
\citep[see][for more detailed reviews]{1985ARA&A..23..169D,Nindos08,Pick08}.

The most usual incoherent mechanisms are free-free emission and
gyroemission.  The former is generated when free electrons interact
with ambient ions via Coulomb forces and is therefore
ubiquitous. Radio emission  is produced when the relevant particle
populations are part of the same thermal distribution and provides
diagnostics of temperature, density, and (to some extent) magnetic
field.

Gyroemission is produced when free electrons are accelerated in a
magnetic  field because they experience the magnetic part of the
Lorentz force.  In the literature gyroemission is referred to as
gyroresonance emission when nonrelativistic electrons are involved and
gyrosynchrotron when mildly relativistic electrons are involved. In
the former case, the  emission comes from hot ($T_e \sim 10^6$ K)
thermal plasma interacting with  magnetic fields in excess of 100
G. It is very sensitive to magnetic field strength and orientation
\citep[e.g.][]{Zheleznyakov70} and generates microwave  emission at low
harmonics of the gyrofrequency above sunspots with strong  magnetic
field. Gyrosynchrotron is the basic microwave process in flares and is
produced by thermal or nonthermal energetic electrons (tens of keV to
several MeV) at harmonics ~10-100 of the gyrofrequency. In addition to
the magnetic field strength and orientation, the emission is also
sensitive to the properties of the radiating electrons
\citep[e.g.][]{Fleishman03b,Fleishman03a}.

The most important coherent emission mechanism is the plasma emission
mechanism \citep[e.g.][]{Melrose80}. It involves the formation of
non-equilibrium electron distributions, which excite plasma waves via
nonlinear processes. The plasma waves can be coverted in part to
electromagnetic waves at the plasma frequency ($\nu_{pe}=9\sqrt{n_e}$
kHz, where $n_e$ is the number density in cm$^{-3}$)  and/or its
second harmonic, $2\nu_{pe}$, via scattering off thermal ions or
coupling to ion-acoustic waves. Radio emissions via the plasma
mechanism are called type III bursts when they are associated with
upward-propagating beams of energetic electrons and type II bursts
when the energetic electrons are accelerated in MHD shocks.

Transient plasma emission is often the brightest solar emission and is
typically observed  at decimeter and longer  wavelengths. The
relevant flux densities may exceed 10$^5$ sfu \citep[e.g.][]{Pick05b} while
the associated brightness  temperatures may lie in the range of
10$^9$-10$^{12}$ K \citep[][]{Sainthilaire13}. On the other hand the
large-scale quiet free-free flux density may  range from less than 1 sfu to
more than 100 sfu ($T_b$ from $\leq 10^6$ to  $\sim$10$^4$ K,
respectively) as we move from low to high frequencies in the SKA
frequency range \citep[e.g.][]{Benz09}. Furthermore, gyroresonance-associated
flux densities may reach 10 sfu ($T_b \sim 10^6$ K) while the gyrosynchtron
mechanism may yield flux densities in excess of 1000 sfu ($T_b > 10^7$ K) at
short microwaves \citep[e.g.][]{Nita02}, and as weak as 0.01 sfu ($T_b \sim
100$ K) in the meter-wave band \citep[e.g.][]{Bastian01,Tun13}.

The above discussion indicates that high dynamic range imaging
is needed to map both the features associated with plasma emission
and those associated with other mechanisms. Probably the most
extreme requirements are posed by the need to map both the plasma
and free-free emissions in CMEs (see Section 9.1) where a dynamic
range of at least 10$^4$ is needed. The dynamic range of SKA1 images
will certainly exceed that value because one of its precursors
(i.e. the MWA) is already  able to exceed it (Mondal et
al. 2018, in preparation). SKA1 images with dynamic range of about
10$^5$ appear to be feasible although one cannot  guarantee values
of about $10^6$ or higher.

The weakest solar radio emissions are associated with tiny transient
activity originated from either free-free or gyrosynchrotron mechanism
(see Section 7). In the literature the weakest among such events have
been presented by \citet{Krucker97} (flux densities down to 0.002 sfu in
events detected at 15 GHz). It is not known whether there exists a
lower limit for the flux density of such weak events but SKA1 unprecedented
sensitivities (see Table 2) will help us detect them down to
levels not achievable before.

\section{Plasma diagnostics of the nonflaring Sun}

In the weak magnetic field regions of the non-flaring Sun, the radio emission
comes from the free-free mechanism. The radiation is formed in local
thermodynamic equilibrium (LTE) and the source function is Planckian. For
radio frequencies the Rayleigh-Jeans approximation is valid and the observed
intensity is proportional to the kinetic temperature of the emitting plasma
for optically thick sources (in stark contrast to the situation in optical
and UV wavelengths).

\begin{figure}
\centering
\includegraphics[scale=0.40]{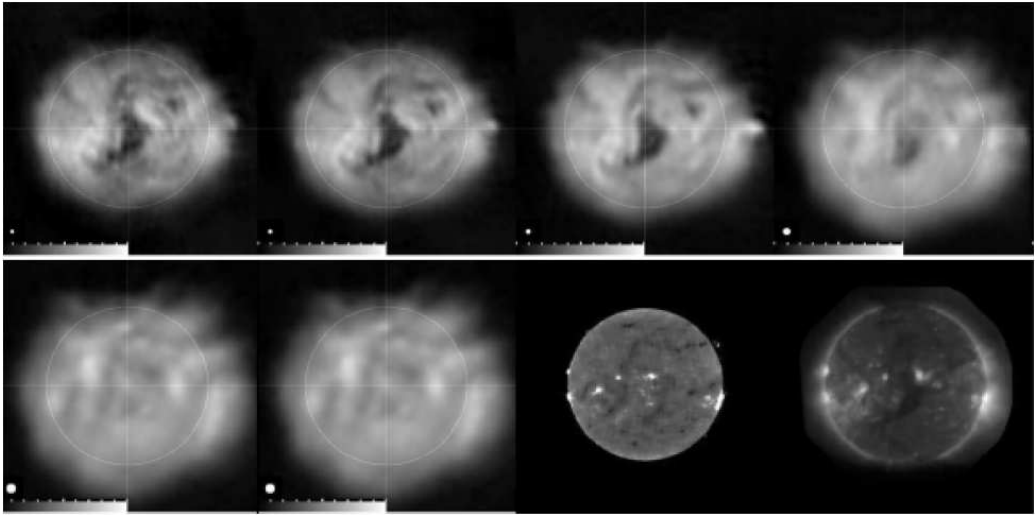}
\caption{Images of the Sun on 2004 June 27. From left to right and from top
to bottom: NRH images at 432, 410, 327, 236, 164, and 150 MHz, NoRH image
at 17 GHz and soft X-rays image from SXI on board GOES12.
After \citet{Mercier09}. Reproduced by permission of the 
AAS.}
\label{fig3}
\end{figure}

The highest frequencies that will be accessible with SKA1 ($3 \lesssim
\nu  < 15$ GHz) are optically thin (an example appears in the 17~GHz
image  of Figure~\ref{fig3}) and we see down below the transition
region (TR): the higher the frequency the deeper we see. At such high
frequencies the only regions above the chromosphere that may  become
optically thick are associated with sunspots of strong magnetic field
where gyroresonance  absorption could become significant. At
frequencies between 1 and  3 GHz the non-flaring corona is
optically thin except  over active regions where hot, dense loops
may render it  optically thick to free-free
\citep[e.g.][]{White99}. At decimeter and longer wavelengths the
emission is optically thick and TR/coronal brightness temperatures are
expected.  Examples of images, produced using rotational synthesis
imaging, at such  low frequencies (obtained with
the NRH at 150-432  MHz) appear in Figure~\ref{fig3}.  The figure
shows that the similarity of the soft X-ray image with the NRH images
decreases as frequency decreases.  This cannot be attributed to a
spatial resolution effect only \citep[][]{Kontar17}.  Apart from
refraction effects that are at work in radio
\citep[e.g.][]{Alissandrakis94,Shibasaki11}, the optical depth of the
radio emission is appreciable.  Therefore these radio  images probe
higher layers of the corona, and lower-lying X-ray-emitting
structures are obscured by overlying dense material. The same
arguments  explain the little resemblance between the NRH images and
the 17 GHz image.

Using the SKA1 frequency bands we will be able to sample, with
unprecedented angular and spectral resolution, the thermal state of
optically thick structures (e.g. the quiet Sun, coronal holes, active
regions, and filaments)  at heights above the chromosphere.  We also
note that there is a lack of imaging of the Sun at frequencies from
0.45 GHz to 1 GHz.  With SKA1-MID Bands 1 and 2 (frequencies from 0.35
to 1.76 GHz) we will be able to image that frequency range
(MUSER, however, may obtain images at that frequency range
before  SKA1).

\section{Coronal magnetography}

Due to inherent difficulties in measuring the Zeeman effect
in the corona \citep{2004ApJ...613L.177L},
radio methods are of primary importance in obtaining quantitative
information about its magnetic field.

\begin{figure}
\centering
\includegraphics[scale=0.50]{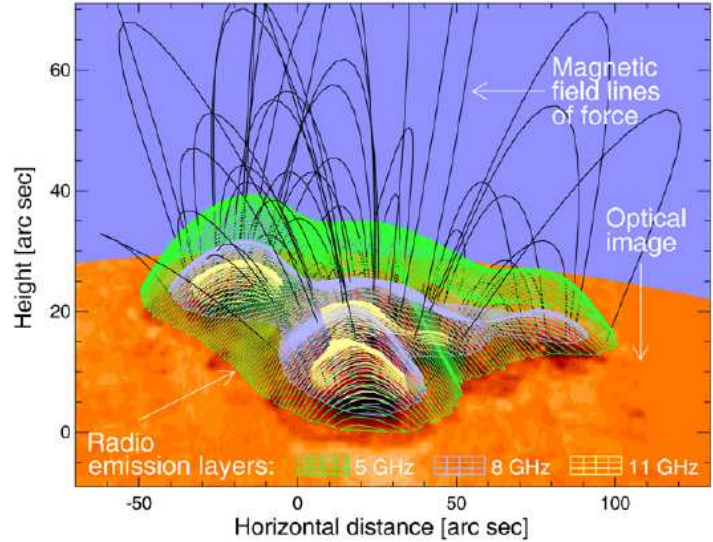}
\caption{Iso-Gauss surfaces corresponding to the third harmonic of the
gyrofrequency at 5, 8 and 11 GHz, together with lines of force of the
extrapolated magnetic field for an active region. From \citet{Lee07}.
Reproduced with permission {\textcopyright}Springer 
Nature.}
\label{fig4}
\end{figure}

\subsection{Coronal magnetic field from gyroresonance emission}

Gyroresonance  emission is a valuable tool for magnetic field
measurements above sunspots. There are several publications on the
subject particularly after the first modeling of high-resolution
observations by \citet{Alissandrakis80}. More recent publications have
been reviewed by \citet{White97} and \citet{Lee07}.

The gyroresonance emission observed at a given frequency originates
from a  narrow iso-gauss layer where the frequency matches a low
harmonic of the  gyrofrequency (see Figure~\ref{fig4}).
The gyroresonance opacity \citep[][]{Zheleznyakov62}
is much higher in the $x$-mode than in the $o$-mode,
and at the second harmonic than at the third.
Above sunspots, the third harmonic is opaque in the $x$-mode but not in the
$o$-mode, and the second harmonic is opaque in both modes. Emission
from the fundamental  is not observed, both  because it is obscured by
the overlying second harmonic  layer and because of propagation
effects, while emission at the fourth harmonic can appear at long
centimeter wavelengths.

For low photospheric magnetic fields or/and high frequencies both the
third  and second harmonic layers are below the TR and  no strong
gyroresonance  emission is expected. The third harmonic layer enters
into the TR for appropriately higher values of the field or lower
frequencies and yields strong, almost 100\% polarized, emission
\citep[e.g.][]{Shibasaki94,Nindos00}. The second harmonic
layer enters the TR for even higher field intensities or lower
frequencies and then the polarization drops.  Therefore the brightness
temperature spectra of both total intensity ($I$) and circular
polarization ($V$) show a rapid rise at the wavelength where the third
harmonic enters into the TR; the magnetic field at the base of the TR
can be estimated from the extrapolation of the $V$ to zero and the
equation $\nu = s \nu_{ce}$ with harmonic number $s = 3$. The potential
of this technique has been demonstrated in several publications that
use OVSA \citep[e.g.][]{Gary87,Gary94} and RATAN-600 spectral data
\citep[e.g.][]{Akhmedov86,Korzhavin10}. Note also that
such measurements are now routinely available from RATAN-600 data and
can be found at \url{http://www.sao.ru/hq/sun}.

SKA1-MID Band 5 (4-15 GHz) observations will provide a brightness
temperature  spectrum along each line of sight through the
gyroresonance source, thereby  enabling the assembly of a map of the
magnetic field at the base of the TR. Given the fact that when a
harmonic layer is opaque the observed brightness temperature is equal
to the local electron temperature, the 3D structure of the coronal
magnetic field could also be constrained \citep[e.g.][]{Tun11} because
the SKA1-MID's spectroscopic imaging observations could yield the
variation of the magnetic field strength as a function of temperature
along each line of sight through the gyroresonance source, provided
that one identifies the harmonic(s) that contribute to the
gyroresonance emission at each frequency.  The dense spectral coverage
that will be provided by SKA1 will yield complete sampling of a large
part of the TR/coronal volume over sunspots, and therefore continuous
magnetic field strength coverage.

\subsection{Coronal magnetic field from propagation effects}

When radio emission passes through a region where the longitudinal
component of the magnetic field changes sign, its sense of circular
polarization reverses when the coupling (which depends on the magnetic
field strength and the gradient of the angle between the line of sight and
the field along the ray path, as well as the electron density and the
frequency of observation)  between the $x$- and $o$-modes is weak and
does not change when the coupling  is strong \citep[e.g.][]{Bandiera82}.
For intermediate ``critical'' coupling the polarization becomes
linear. The  line that delineates the reversal in the sense of
circular polarization ($V=0$) is known as ``depolarization strip''.

In depolarization strips one can estimate the magnetic field under
some assumption about the electron density \citep[e.g.][]{Kundu84}.
The extension of this method for cases of partial depolarization
allows for the computation of a coronal magnetogram over the
quasi-transverse layer (i.e. the region where the magnetic
field is almost perpendicular to the direction of propagation of the
radiation), provided that the intrinsic polarization of  the waves
is known, for example by using polarization data before the mode
coupling occurs \citep[see][for details]{Ryabov99,Ryabov05}.

Polarization inversion provides diagnostics about the coronal magnetic
field  at heights of about 0.05-0.4$R_{\odot}$, and is independent of
the emission mechamism. Therefore all SKA1-MID Bands could be
used to  exploit the potential of the method: their dense spectral
coverage will allow to locate the coronal regions where the field is
perpendicular to the line of sight and to derive information about the
magnetic field well above the formation height of the microwave
emission.

\subsection{Magnetic field from free-free emission}

The weak polarization of the free-free emission in the sense  of the
$x$-mode can be used to determine
the longitudinal component, $B_l$, of the magnetic field:
for an optically thin slab above a uniform background
it is easy to show that $B_l$ is proportional to the degree
of circular polarization, $\rho$, and the frequency.
Things are more complicated in the general case, where
physical conditions vary with height.
When spectral observations are available,
one can use the approximate expression \citep{Bogod80,Grebinskij00}
to determine $B_l$:
\begin{equation}
B_l \approx 107 \frac{\rho (\%)}{n \lambda},
\end{equation}
where $B_l$ is in G, $\lambda$ is the wavelength in cm,
and $n = - \ln T_b / \ln \nu$ the spectral index.
This expression allows for temperature variations
in the  radiation-formation region and
its validity is not limited to the  optically thin case,
although it implicitly assumes constant magnetic field.
The combination of SKA1-MID's unprecedented sensitivity and high
frequency resolution will allow, for the first time, to constrain the
longitudinal  component of the coronal magnetic field to a few Gauss.

An extension of the above method could be used to provide estimates of
the magnetic field strength of undisturbed streamers at higher
altitudes (heliocentric distances of 1.5$R_{\odot}$ to
2.5$R_{\odot}$) by measuring the degree of circular polarization of
the associated free-free emission using SKA1-LOW observations
\citep[see][for details]{Sastry09,RameshSastry10}.

\section{Coronal heating}

Coronal heating remains one of the major unsolved problem in solar physics.
The proposed mechanisms to heat the corona
\citep[e.g. see][for reviews]{Klimchuk06,Demoortel15}
are divided into two broad classes:
waves and small-scale reconnection events, also known as
nanoflares. Acoustic wave heating has been a favorite for the non-magnetic
chromosphere, but it appears that the appropriate waves
do not  carry enough energy \citep[][]{Fossum05}.
Other types of waves, such as Alfven waves in the
presence of turbulence \citep[][]{Vanballegooijen11} could be an alternative.

\begin{figure}[t]
\centering
\includegraphics[scale=0.70]{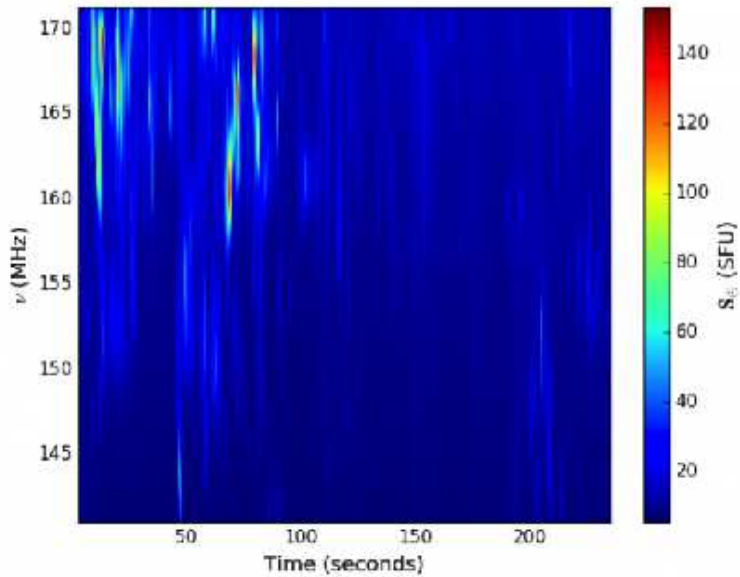}
\caption{MWA dynamic spectrum with numerous nonthermal, weak (0.6-307 sfu)
narrow-band (4-5 MHz), short-lived (1-2 s) bursts. This figure has been
constructed with data presented in \citet{Suresh17}.}
\label{fig5}
\end{figure}

There have been several efforts to search radio data for small-scale
transient events that could potentially contribute to coronal
heating. These efforts are partially inspired by the sensitivity of the
radio domain to small  populations of nonthermal electrons that could
arise in the course of  reconnection. At microwaves such studies
\citep[e.g.][]{White95,Krucker97,Gary97,Nindos99,Benz99}
have shown that tiny transient activity may occur both within
and away from ARs and that in several cases the events show
flare-like properties being either thermal or nonthermal.
At lower frequencies short duration small isolated bursts have also been reported
\citep[e.g.][-see Figure \ref{fig5}]{1986ApJ...308..436K,Ramesh10,Sainthilaire13,Suresh17}.


Several authors \citep[e.g.][]{1984ApJ...283..421L,Gary99,2000ApJ...535.1047A,2002ApJ...572.1048A}
have computed histograms of the energy distribution
in impulsive events using X-ray and EUV data.
However, the low energy part of flaring events
is poorly known, hence nanoflare heating model \citep{1988ApJ...330..474P}
evades observational confirmation.

SKA1 could provide significant insights into the coronal
heating problem because: (1) it will provide detailed time series of
the temperature, density, and magnetic field in various coronal
structures (see Sections 5 and 6) from which the rate of energy
storage into the corona could be estimated as a function of position
and time and then checked against wave heating model predictions,
(2) due to its unprecedented sensitivity, angular, temporal and spectral
resolution, as well as wide field of view, SKA1 will be able to detect
weak, small, short-lived, narrow-band transient events down to levels
not achievable earlier, over large parts of the solar disk.
The computation of the occurrence rate and energy budget of these events
will help us derive more accurate constraints than ever before
and evaluate their role in coronal heating.

\section{The physics of flares}

Through magnetic reconnection, flares release large amounts of magnetic
energy  stored in the corona which accelerates particles
and heats extended layers of  the solar atmosphere.
Accelerated particles emit at radio wavelengths, X-rays, and in some cases gamma-rays \citep[see][for a review]{2011SSRv..159..107H}.
The hard X-rays typically arise from the interaction of electrons
with energies from a few tens of keV to a few hundreds of keV
with chromospheric plasma via the nonthermal free-free mechanism.
On the other hand, radio emission may arise either from suprathermal electrons
that produce coherent radio emission due to plasma instabilities or from
mildly relativistic electrons that emit nonthermal gyrosynchrotron
radiation. X-ray and radio observations complement each other
\citep[e.g.][]{White11} and both are essential to study particle
acceleration and transport. Radio observations provide the advantage
that not only are they sensitive to anisotropies in the electron
distribution function, but also uniquely sensitive to the magnetic
field.

\subsection{Electron acceleration}

Flare energy release and particle acceleration appear to be
spatially fragmented processes, where the acceleration region
may not coincide with the region of magnetic energy
release \citep[e.g. see the reviews by][]{Bastian97,Zharkova-ssr11}.
Furthermore, the energy partition \citep{2012ApJ...759...71E,2017ApJ...836...17A}
and the efficiency of  the particle acceleration may vary from event to event,
from purely thermal \citep[e.g.][]{Gary89} to acceleration-dominated
\citep[e.g.][]{Krucker10,Fleishman11}.

The high sensitivity of radio emission to nonthermal electrons makes it
possible to probe rather tenuous electron acceleration sites that could
not be detected by hard X-rays which require higher densities
as observed in some events \citep{2008ApJ...673..576X,2011ApJ...730L..22K}.
Radio diagnostics of the electron acceleration region include
(1) decimetric type III bursts,
(2) decimetric narrow-band spikes,
and (3) microwave narrow-band gyrosynchrotron bursts.

Type III bursts are plasma emissions resulted from the propagation of
beams of energetic electrons along open field lines
\citep[see][for reviews]{1985srph.book..289S,Reid14a}.
Of particular interest are the decimetric type III bursts  that usually (but not always)
occur in the 400 MHz to 1 GHz frequency range which corresponds to densities of
$2-12 \times 10^9$ cm$^{-3}$; i.e. to the  heights where flare energy
release occurs \citep[e.g.][]{Benz04}. \citet{Reid11,Reid14b} combined
hard X-ray spectroscopic data and radio dynamic spectra of type III bursts
to find that the acceleration region is located well above soft X-ray
flaring loop tops at heights between 25 to 200 Mm. This result is
consistent with earlier findings by \citet{Aschwanden97}.


The potential of exploiting decimetric and metric type III bursts for
the  diagnosis of the electron acceleration site is hampered by the
fact that  relevant multi-frequency imaging observations are rare and
only at a few discrete frequencies
\citep[e.g.][]{Aurass97,Paesold01,Alissandrakis15}. 
Probably the only imaging spectroscopic observation of a type III 
burst has been reported by \citet{Chen13} who found that the radio data 
indicated that the energy release showed fragmentation in space and time, 
suggesting a bursty reconnection scenario.

Among the coherent transient emissions at decimetric wavelengths,
spikes are expected to show the closest relation to the electron
acceleration region because (1) they show the highest correlation with
hard X-rays \citep[e.g.][]{Benz86,Aschwanden92}
and (2) they occur at the rise phase of the flare \citep[][]{Slottje78}.
However, studies of the locations of decimetric  spike bursts and hard X-ray sources
suggest the presence of a secondary acceleration region high up in
the corona, away from the flaring site \citep[][]{Benz02,Khan06,Battaglia09}.

\citet{Fleishman11,Fleishman13,Fleishman16} have shown
that narrow-band microwave  gyrosynchrotron radiation
which is emitted directly from the accelaration  region can be detected
when the trapped electron population is negligible.
This emission distinguishes itself from the usual broadband gyrosynchrotron
emission due to the steep energy spectrum of electrons accelerated
in a rather uniform source.

\subsection{Electron transport}

Electron transport during flares is usually discussed in relation to
the dynamics of both the radio and associated hard X-ray emission. The
most widely used model is the ``direct precipitation/trap plus
precipitation'' (DP/TPP) model \citep[e.g. see][as reviews]{Bastian98,Aschw-ssr02}.
Briefly, the magnetic field guides accelerated
electrons with small pitch  angles to the chromosphere, where the
dense plasma stops them \citep[e.g.][for a review]{2011SSRv..159..107H}.
The largest part of their energy heats the ambient chromosphere
while a smaller part is emitted in HXRs via the nonthermal
``thick-target'' free-free mechanism.
Electrons with sufficiently large pitch angles are trapped in the flaring loop
and produce gyrosynchrotron emission. 
Eventually, they are scattered into the loss cone through wave-particle 
interactions or Coulomb collisions and precipitate into the chromosphere,
producing additional hard X-ray emission.
Microwave gyrosynchrotron emission does not come
exclusively from trapped  electrons;
in several studies microwave emission has been reported
from the precipitating electrons \citep[e.g.][]{Kundub01,Lee02}
or from electrons that are efficiently scattered \citep{2018A&A...610A...6M}.

The observed signatures of gyrosynchrotron radiation are sensitive
to anisotropies in the electron distribution function \citep[e.g.][]{Lee00}.
However, it is not always straightforward to deduce whether
the anisotropies stem from the injection/acceleration of the electrons
or from transport effects \citep[e.g.][]{Melnikov02} and it appears
that acceleration and transport are interwined in some cases
\citep[e.g.][]{Bastian07}.

\subsection{Magnetic field diagnostics in flaring loops}

\begin{figure}
\centering
\includegraphics[scale=0.50]{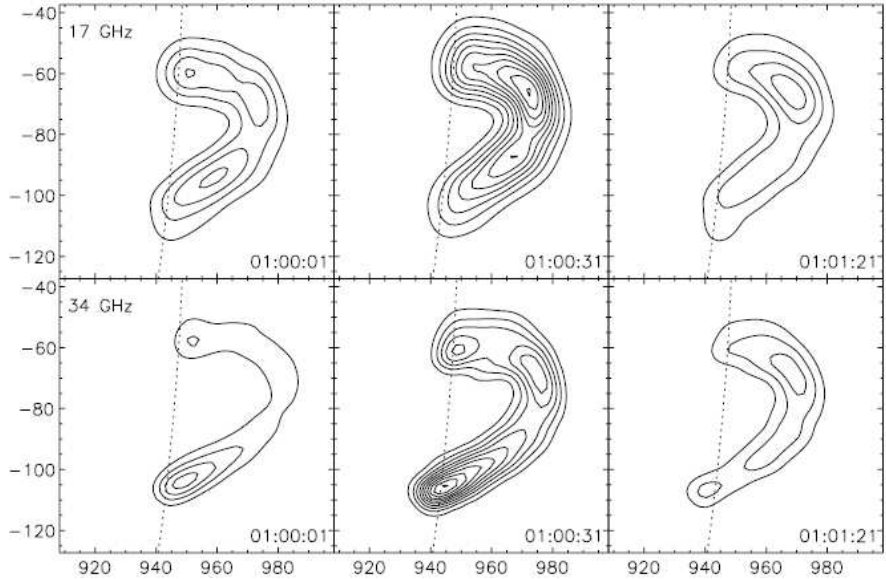}
\caption{A large flaring loop observed at 17 and 34 GHz with the NoRH.
Contours are at 10\% intervals of 550 and $170 \times 10^6$ K at 17 and
34 GHz, respectively. From \citet{Tzatzakis08}. Reproduced with 
permission {\textcopyright}Springer Nature.}
\label{Fig7}
\end{figure}

The spectral and spatial structure of microwave gyrosynchrotron
emission from flaring loops has a strong dependence on the magnetic
field and the properties of energetic electrons. The spectral peak
divides the spectrum into an optically thick low frequency part and an
optically thin high frequency part. In a flaring loop the magnetic
field is strongest near the footpoints and decreases toward the loop
top.  Therefore, at a given frequency the higher the energy of the
electrons, the  closer to the flaring loop top will they
emit. Consequently, for a homogeneous  isotropic distribution of
energetic electrons, at high frequencies we expect  to obtain
optically thin emission primarily from the footpoints of the flaring
loop \citep[e.g.][]{Alissandrakis93} while more extended optically
thick  sources should appear with the decrease of the  frequency of
observation \citep[e.g.][]{Wang94}.
Anisotropies or/and inhomogeneities in the  properties of the electrons
may modify the above picture, for example by  the development of optically
thin loop top sources due to high concentration of energetic electrons there
\citep[][]{Melnikov02} \citep[see also][]{Kundu01,White02,Tzatzakis08}.
An example of the diversity of gyrosynchrotron source morphologies is
presented in Figure~\ref{Fig7}.

The above discussion indicates that the use of gyrosynchrotron
emission for diagnostics of the magnetic field is not as
straightforward as in the case of gyroresonance.
For reliable diagnostics one would need to combine detailed modeling
of the gyrosynchrotron emission
\citep[e.g.][]{Klein84,Alissandrakis84,Bastian98,Fleishman10,Simoes10,Kuznetsov11,2015ApJ...799..236N}
with spectroscopic imaging observations.
Thus any information on the magnetic field comes as a result of modeling.
In spite of the difficulties, modeling of individual events observed either by the VLA
and OVSA \citep{Nindos-white00}  or by the NoRH and the Nobeyama
Polarimeter \citep[e.g.][]{Kundu04,Tzatzakis08,Kuznetsov15} have yielded
magnetic field strengths ranging between 1700 G at the footpoints
and 200 G at the loop top.

\subsection{SKA's contribution to flare research}

The discussion in Sections 8.1-8.3 indicates that despite the
significant recent progress, several fundamental questions about the
physics of flares  remain unanswered.
These questions include: (1) The location of  electron acceleration,
its magnetic configuration, and its relation to the  site of energy release.
(2) The mechanism(s) responsible for electron  acceleration
and the conditions under which they operate.
Several acceleration models  have been proposed
\citep[e.g. see][for reviews]{Zharkova-ssr11,Klein17} but all of  them
are in need of reliable
observational inputs. (3) The accurate  determination of the
distribution function of the energetic electrons and its evolution in
space and time. (4) The factors that determine the efficiency with
which magnetic energy is converted into energy of nonthermal
particles.  (5) The physical processes responsible for  the transport
of accelerated electrons and how one can disentangle
injection/acceleration from transport effects.

The unique capabilities of SKA1 have the potential to provide
significant advances in addressing the above questions. First of all,
SKA1 will provide the means to perform coronal magnetic field
measurements before,  during, and after flares (see  Section 6), thus
allowing the determination of the magnetic free energy that  becomes
available. The relevant results will be superior from the ones
provided by nonlinear force-free extrapolations of the photospheric
field because the latter provide non-unique solutions
\citep[e.g.][]{DeRosa09}, rely on field measurements  in layers where
the field is not force-free \citep[][]{Metcalf95}  and their cadence
is not sufficient to capture the impulsive  evolution of energy
release and particle acceleration. Spectroscopic imaging observations
will  allow to establish the location of  electron acceleration and to
provide important constrains on the relevant acceleration
mechanisms by revealing, in unprecedented detail, the properties of
the  bursts that are believed to be intimately  related to the
acceleration process, and most probably by discovering new relevant
diagnostics.  The combination of forward fitting techniques with
high-cadence spatially  resolved spectra of the gyrosynchrotron
emission will allow to constrain both the properties of the electron
distribution function and the magnetic field in the flaring volume
more accurately than ever before.

\section{Large-scale transient activity and space weather}

\subsection{Coronal mass ejections}

All CME constituents emit radio radiation
\citep[see][for a review]{Gopal11}.
Eruptive prominences that later will evolve into
the cores of white-light CMEs are observed at microwaves via the
thermal  optically thick free-free emission that they  emit
\citep[e.g.][]{Hanaoka94,Gopalswamy96,Gopalswamy03,Grechnev06,Gopalswamy13}.
These observations complement
the usual  H$\alpha$ observations and have the advantage that the
continuum free-free emission can be detected even when the material is
heated to high temperatures that make it undetectable in
H$\alpha$. Higher up, one anticipates production of optically thin
free-free emission from the whole CME that could be detected at lower
frequencies. However such observations are rather rare
\citep[e.g.][]{Sheridan78,Gopalswamy92,Kathiravan02,Ramesh03}
because the relevant emission is weak (due to the  high
temperatures and low densities that usually prevail in CMEs) and often
outshined by the stronger nonthermal emissions. When detected, the free-free
emission from CMEs provides an alternative way to estimate their masses
\citep[][]{Gopalswamy92}.

\begin{figure}
\centering
\includegraphics[scale=1.00]{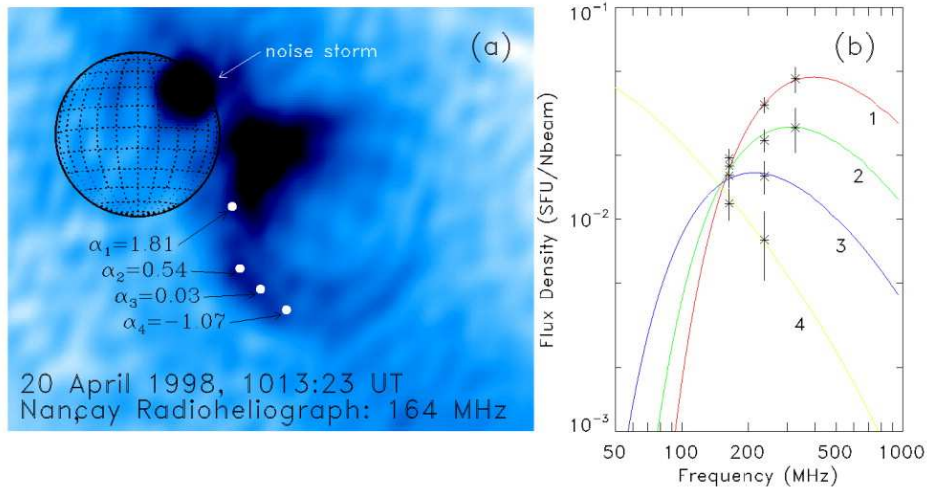}
\caption{(a) Image of a CME at 164 MHz computed from NRH data. The spectral
index estimated at the locations of the four white dots is also marked. (b)
Flux spectra estimated at the points marked in (a) accompanied by model
gyrosynchrotron spectra. From \citet{Bastian01}. Reproduced
by permission of the AAS.}
\label{fig8}
\end{figure}

At radio dynamic spectra, CME-related emissions include type II bursts
(see Section 9.2)  as well as stationary and moving type IV bursts. The
stationary type IV bursts emanate from electrons trapped in
post-eruption  arcades behind CMEs while the moving ones correspond to
CME-related  outward-moving material. Stationary type IV bursts should
arise from  plasma emission mechanism, while plasma emission
\citep[e.g][]{Duncan81,Stewart82,Gopal89,Klein02,Ramesh13,Hariharan16},
and more rarely gyrosynchrotron emission from nonthermal electrons
\citep[][]{Gopal87,Bastian01,Maia07,Tun13,Bain14,Carley17} have beeen
invoked for the interpretation of the properties of moving type IV bursts.
The detection of gyrosynchrotron emission from some CMEs
(see Figure~\ref{fig8} for an example),
is important because, if combined with modeling (see Section 8.3),
it could  provide estimates about the CME magnetic field; in the publications
cited above, CME magnetic fields from 15 to 0.1 G have been reported at 
heights from $1.3R_{\odot}$ to $2.8R_{\odot}$.

Radio imaging observations of the early development of CMEs, albeit at
few frequencies only, have  revealed a large diversity of magnetic
configurations that produce energetic electrons
\citep[e.g.][]{Klein01,Pohjolainen01,Maia03,Pick05a,Pick06,Demoulin12,Carley16}.
Such observations are a necessary ingredient for any validation of CME
initiation models because nonthermal radio emissions could highlight
possible locations of magnetic reconnection.

The potential of radio observations to CME research has not been fully
exploited because of their lack of spectroscopic imaging capabilities
and their inadequate sensitivity. SKA1 capabilities promise to change
that situation. First of all, SKA1's high angular, spectral, and
temporal resolution will provide, for the first time, a comprehensive
picture of CMEs over extended frequency ranges. Since plasma emission
probes the plasma frequency  level or/and its second harmonic the
resulted radio morphologies may be different from the white-light CME
morphologies, therefore for non-incremental  progress, one should be
able to image both the thermal free-free emission and  the much
stronger plasma emission simultaneously. For this task
a dynamic range of at least 10$^4$ is required (see the discussion
in Section 4) which should be well within SKA1's capabilities.
We also note that according to \citet{Bastiangary97}
the optimum frequency range for the detection of free-free
emission from CMEs is between $0.2-2$ GHz (i.e. within SKA1's frequency
range). 

SKA1-LOW will be ideal to probe the early development of CMEs
(including earth-directed ones whose detection by coronagraphs at L1
might be difficult at times), study the flare-CME relationship, and
determine the sites of CME-related electron acceleration. SKA1-LOW's
wide  field of view will allow to study the effects of CMEs on the
surrounding plasma and their evolution  into interplanetary
disturbances. Finally, the spectroscopic imaging capabilities of the
instrument combined with its high sensitivity also promise more
frequent detections of gyrosynchrotron emission from CMEs, and
consequently the development of a statistically significant sample of
CME magnetic field estimates.

SKA1 will also play a key role in studying CMEs in the interplanetary
medium. Along with the more traditional interplanetary scintillation
studies, heliospheric science with the SKA is also  expected to
include studies of the Faraday rotation of linearly polarized
background radiation due to the magnetized solar wind and CME
plasma. In this article we restrict ourselves only to a discussion of
the solar physics questions which can be addressed with the SKA1. While 
important and interesting, the heliospheric physics to be pursued by the SKA1 
lies outside the scope of this article. Interested readers are referred to the
review by \citet{Nakariakov15} for a short discussion of these topics.

\subsection{Coronal shocks}

Type II bursts are plasma emissions resulting from the propagation of
MHD shocks in the corona; the emission appears as narrow-band lanes at
the plasma frequency or/and its second harmonic
\citep[see the reviews by][]{Nelson85,Vrsnak08,Pick08,Nindos08}.
Although interplanetary type II bursts are generated exclusively by
CME-driven shocks \citep{Gopalswamy06}, the combination of radio imaging
observations with coronagraphic, EUV, and soft X-ray data show that
coronal shocks could appear close to either the leading edge
\citep[e.g.][]{Maia00,Ramesh12} or the flanks of CMEs
\citep[e.g.][]{Cho07,Zucca14}, ahead of erupting flux
ropes \citep[][see Figure 8]{Bain12,Zimovets15},
above
expanding loops \citep[][]{Klein99,Dauphin06}, in
conjunction with jets that erupt in the course of CMEs \citep[][]{Zucca-etal14},
and ahead of EUV bubbles \citep[][]{Kouloumvakos14}.
CME deflections \citep{2016ApJ...823....5P} and streamer-CME interactions
\citep[][]{Kong12,Eselevich15} in the low corona could also play a
role in the generation of coronal shocks. Some coronal shocks are
flare-related \citep{Magdalenic08,Magdalenic10,Magdalenic12,Nindos11,Kumar16}
because either there was no CME or the timing of the CME did not match the
timing of the type II burst.

\begin{figure}[ht!]
\centering
\includegraphics[width=\textwidth]{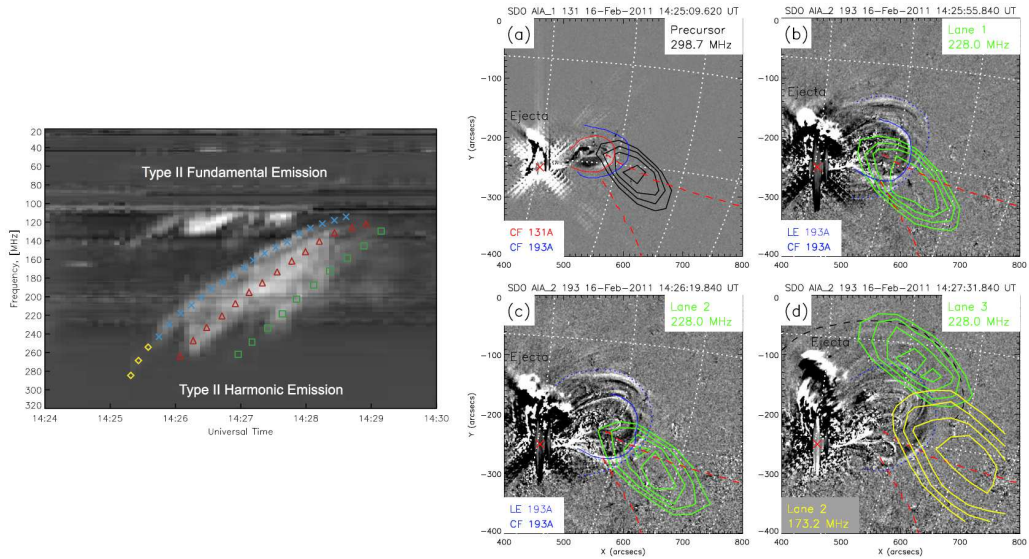}
\caption{Left: a complex type II burst observed by the ARTEMIS IV radio
spectrograph. Yellow diamonds mark precursor activity while the blue,
red, and green symbols mark the central frequencies of the three emission
lanes of the harmonic emission. Right: Contour plots of precursor (a) and
type II (b-d) second harmonic emission of the event in the ARTEMIS dynamic
spectrum. The background consists of AIA 131 \AA\ difference images.
After \citet{Zimovets15}. Reproduced with permission 
{\textcopyright}Elsevier.}
\label{fig9}
\end{figure}

The above discussion indicates that coronal shocks could be produced
by a variety of drivers: CMEs, small-scale ejecta associated with
flares or flare pressure pulses. Some authors (e.g. Gopalswamy et al.
2005, 2009) have concluded that all metric type II bursts are
generated by CME-driven shocks and the popularity of this
interpretation is reinforced by the high association between metric
type II bursts and EUV waves, the latter being driven by the lateral expansion
of CMEs \citep[e.g.][]{Patsourakos12}. However, it is fair
to say that for a few events a flare origin cannot be excluded.

Progress in the study of coronal shocks has been limited by the
inadequate angular and spectral resolution as well as sensitivity of
radio imaging  instruments and by the fact that the coronal shock
emission that they  detect may come from heights occulted in
coronagraphs. SKA1-LOW's specifications promise significant advances
in our understanding of coronal shocks. With its superior
angular-spectral resolution and dynamic range, the instrument will be
able to image both  the CME and shock radio emission  simultaneously
which (combined with observations in the EUV and soft X-rays) will
help us identify unambiguously the role of each eruption constituent
to the development of  the shock from its earliest stages. SKA1-LOW's
wide field of view will also help us monitor the shock propagation and
evolution in the  corona, well above the heights obscured by the
occulting disk of coronagraphs.  Furthermore, the fine structure of
type II bursts \citep[e.g.][]{Mann05,Carley13,Mann18} will be recorded with
unprecedented detail which is an essential step to constrain shock
inhomogeneities and turbulence as well as the role of shocks in particle
acceleration.

\subsection{Origin of solar energetic particles}

Solar energetic particle (SEP) events are important for both the study
of particle acceleration and space weather
\citep[e.g.][]{Reames99,Gopal08,Vainio09,Valtonen11}. Since the
work by \citet{Reames99} SEP events are usually classified into two
groups: impulsive or  gradual depending on the strength and duration
of the parent soft X-ray flare, their association with type III or
type II/IV bursts, charge states and abundances of the energetic
particles, and the absence or presence of a CME. Impulsive events is
thought to originate from particle acceleration induced by
flare-related  magnetic reconnection while the accelearation of
gradual-event particles is thought to originate from CME-driven
interplanetary or/and coronal shocks.

The above dichotomy has been challenged by several particle and
radio/hard X-ray studies
\citep[e.g.][]{Klein-chupp99,Klein14,Laitinen00,Klein-trot01,Cane02,Klein05,Kouloumvakos15}
which indicate
that it is difficult to establish a clear distinction between
CME-associated and flare-associated SEP events. After all,
reconnection, for example, could occur not only during the impulsive
phase of flares but also in current sheets that are formed behind CMEs
while acceleration by shocks generated by the interaction of
reconnection outflows with loops in a cusp-shaped configuration
might also be possible \citep[e.g.][]{Aurass02,Aurass04,Mann09,Warmuth09,Chen13}.

Sections 8 and 9.1-9.2 indicate that SKA will probe in an unprecedented way
the  diverse agents of particle acceleration in the low and middle
corona and  therefore we expect that its observations will provide key
inputs that will help shed light on the controversies related to the
origin of SEPs.

\section{Synergistic activities}

SKA's construction overlaps with the operation and/or development of
important  new-generation ground-based and space-borne solar
instruments.

Ground-based facilities include Big Bear Solar Observatory's Goode Solar Telescope
\citep[in operation since 2009;][]{Goode12},
the Daniel K. Inouye Solar Telescope
\citep[][under construction with a planned completion date of 2019]{Tritschler15}
and the European Solar  Telescope
\citep[][under construction with a planned completion date of 2020]{Matthews16}.
These instruments observe (or could) observe the photosphere
and chromosphere in optical or/and near-infrared wavelengths
with unprecedented sensitivity (owing to their large apertures),
angular resolution ($\sim 0.1\arcsec$ using adaptive optics),
and polarization accuracy ($\sim 10^{-4}$ of intensity).
Their science objectives include the study of (1) solar magnetic fields
(their life cycle and their role in the initiation of transient activity),
and (2) the mechanisms of solar  variability.
The SKA will not be able to observe the layers probed by the above
instruments, but coordinated observations will be of major importance
providing a comprehensive picture of all layers of the solar atmosphere,
from the photosphere to the outer corona.

The most important forthcoming solar space missions
are the Parker Solar Probe \citep[PSP,][]{Fox16}
and the Solar Orbiter \citep[SO,][]{Muller13},
which are scheduled for launch in 2018 and 2020, respectively.
Both missions are designed to approach the Sun closer
than ever before (10 and 60$R_{\odot}$, correspondingly)
and carry several instruments for in situ measurements.
As for remote sensing instruments they will both be equipped
with heliospheric imagers while SO will also be equipped
with a vector magnetograph, instruments for EUV imaging and spectroscopy,
a telescope/spectrometer for thermal and nonthermal X-ray emission,
and a coronagraph. SKA1 and SO/PSP data will complement each other.
For example, due to their close approach to the Sun both SO and SPP could
measure SEP properties with minimal influence of transport effects,
while SKA1 will provide information about energetic electron seed populations
and the role of flares and CME-driven shocks in SEP acceleration.
The combination of SKA observations with data from the in situ  radio wave
instruments (available from both SO and PSP) will allow
to track (albeit spatially unresolved) CME-related emissions recorded
by the SKA1 into the interplanetary medium.
Measurements of the magnetic field in the lower solar atmosphere
by SO will complement the SKA1's coronal magnetography capabilities,
while the combination of SKA1 data with SO hard X-ray observations
could provide new opportunities to study flare electron acceleration
low in the solar atmosphere. SO/PSP synergies with the SKA1 will also
have a strong heliospheric component because SKA1 observations
could provide constraints  on turbulence and waves in the solar wind
\citep[see][for details]{Nakariakov15}.
Finally, needless to say that coordinated observations between the SKA1
and currently existing space instruments (e.g. SDO, RHESSI, STEREO, Hinode,
and IRIS) could also be very useful.

SKA1-LOW solar observations will overlap in time with solar-dedicated
radio heliographs that will observe at higher frequencies (NoRH,
MUSER, and Siberian RH). The temporal overlap could be exploited to
partially compensate for the most probable inability to operate the
SKA1-MID and LOW arrays simultaneously. The successful exploitation
and interpretation of several SKA1 observations will benefit from recent developments in
theory \citep[e.g.][]{Zharkova-ssr11,Hannah13,Kontar15,Kontar-perez17}
and simulations of solar radio emissions
\citep[e.g.][]{2011PhPl...18e2903T,Li12,Gordovskyy14,Schmidt14,2014A&A...572A.111R,Schmidt16}.

\section{Summary}

The SKA is a unique radio instrument which promises to transform
practically all branches of astrophysics due to its unique
specifications  which, even after completion of the Phase 1 of its
deployment (i.e. SKA1),  could significantly outperform all other
radio telescopes. Solar physics benefits immensely from the
deployment of the SKA because of the feasibility of solar observations
with both SKA1-LOW and SKA1-MID.
SKA1's unprecedented angular, spectral, and temporal resolution, as
well as sensitivity will provide major new insights into many important
solar physics problems. 

Observations of the non-flaring solar atmosphere
will provide time series of its thermal state which can be
checked against solar-atmosphere heating models.
The detection of numerous weak transient events could
facilitate the derivation of reliable estimates about their
contribution to coronal heating in the framework of the
nanoflare model.

Several radio emission mechanisms/processes are uniquely
sensitive to magnetic field, and one of the most important outcomes
of SKA1 observations will be the direct and indirect measurements
of the magnetic field at heights unaccessible by other instruments.
The measurements can be used both for the computations of magnetic
free energy budgets as well as for the diagnosis of the magnetic field
of active regions, flaring loops, and CMEs.

SKA1 observations have strong potential to
provide a comprehensive view of coherent and incoherent emissions
that are intimately related to electron acceleration,
of gyrosynchrotron emission from precipitating
and trapped electrons in flaring loops, as well as of CMEs, shocks,
and related phenomena.
These observations have the potential to provide major advances
to key solar physics questions about:
(1) the location and magnetic configuration of the electron acceleration site,
(2) the mechanism(s) responsible for particle acceleration,
(3) the flare-CME relationship,
(4) the timing  and evolution of CMEs from the early stages
of development all the way to the outer corona,
(5) the drivers of coronal shocks as well as the locations
and efficiency of electron acceleration by shocks,
and (6) the origin of SEPs.

Above all, as is always the case with new instruments that outperform
their predecessors in a significant way, is the high probability of
new discoveries that cannot be predicted now.  This exciting prospect
is further highlighted by the availability of synergistic activities
between the SKA and the new generation of ground-based and space-borne
solar instruments.

\section{Acknowledgements}

We acknowledge the activities of the members of the SHI SWG
toward the goal to exploit SKA's capabilities for solar physics
research. We thank the referees for their comments which led to
improvement of the paper. EPK was supported by  STFC consolidated
grant ST/P000533/1. DO was partially supported by a grant  from the
Department of Atomic Energy, Government of India, for enabling
Indian  participation in the SKA.

\section*{References}


\end{document}